\documentclass[11pt]{article}
\usepackage[sfdefault]{AlegreyaSans} 
\usepackage{geometry}
\usepackage{setspace}
\usepackage{titlesec}
\PassOptionsToPackage{hyphens}{url}\usepackage{hyperref}
\usepackage{graphicx}
\usepackage{amsmath}
\usepackage{bm}
\usepackage{wrapfig}
\usepackage{xcolor}
\usepackage{authblk}
\usepackage[export]{adjustbox}
\usepackage{subcaption}
\usepackage{tikz} 
\usetikzlibrary{matrix}
\usepackage[numbers,sort&compress]{natbib}
\usepackage{comment}
\usepackage{upgreek}

\titleformat{\section}{\normalfont\fontsize{12}{14}\bfseries\sffamily}{\thesection}{1em}{}
\titleformat{\subsection}{\normalfont\fontsize{11}{13}\bfseries\sffamily}{\thesubsection}{1em}{}
\titleformat{\title}{\normalfont\fontsize{14}{16}\bfseries\sffamily}{}{0em}{}

\title{On-Device Control of Electronic Friction}

\author[1,2]{Zhaokuan Yu}
\author[2]{Jinbo Bian}
\author[3,4]{Jin Wang}
\author[5]{Zonghuiyi Jiang}
\author[2]{Linxin Zhai}
\author[1]{Xin Lu}
\author[5]{Xiaofei Liu}
\author[2,6,7]{Quanshui Zheng$^\ast$}
\author[2]{Zhiping Xu$^\ast$}
\affil[1]{Center for Correlated Matter, School of Physics, Zhejiang University, Hangzhou 310058, China}
\affil[2]{Center for Nano and Micro Mechanics, Applied Mechanics Laboratory, Department of Engineering Mechanics, Tsinghua University, Beijing 100084, China}
\affil[3]{International School for Advanced Studies (SISSA), Via Bonomea 265, Trieste 34136, Italy}
\affil[4]{International Centre for Theoretical Physics (ICTP), Strada Costiera 11, Trieste 34151, Italy}
\affil[5]{State Key Laboratory of Mechanics and Control for Aerospace Structures, Key Laboratory for Intelligent Nano Materials and Devices of the Ministry of Education, Nanjing University of Aeronautics and Astronautics, Nanjing 210016, China}
\affil[6]{Tsinghua Shenzhen International Graduate School, Tsinghua University, Shenzhen 518055, China}
\affil[7]{Institute of Superlubricity Technology, Research Institute of Tsinghua University in Shenzhen, Shenzhen 518057, China}
\affil[$^\ast$]{Corresponding author(s): Zhiping Xu (xuzp@tsinghua.edu.cn), Quanshui Zheng (zhengqs@tsinghua.edu.cn)}

\date{}

\newlength{\tempdima}
\newcommand{\rowname}[1]
{\rotatebox{90}{\makebox[\tempdima][c]{\textbf{#1}}}}

\newcommand\zfig[1]{{\color{violet}#1}}

\usepackage[T1]{fontenc}
\usepackage{array}
\usepackage{booktabs}
\usepackage{float}

\begin{document}

\maketitle

\begin{abstract}
Friction causes mechanical energy dissipation and material degradation in machinery and devices.
While phononic friction is well understood via anharmonic lattice dynamics, the physics of electronic friction remains unclear due to challenges in separating electronic degrees of freedom from phononic ones in experiments and analyzing the non-equilibrium interactions between ionic movement and electronic dynamics in theory.
To tackle this problem, we construct a sliding device featuring 2D crystalline interfaces that possess ultra-smooth and minimally interacting surfaces, achieving the state of structural superlubricity with no wear and minimal friction.
Using electrical and mechanical controls, we tuned the nature of interfacial electronic coupling and charge densities in materials in an \emph{on-device} manner, which allows us to disentangle the electron and phonon contributions to friction.
Our experimental data and theoretical analysis supported by first-principles calculations demonstrate that electronic friction can well surpass phononic contributions and dominate energy dissipation at structural superlubricity contacts.
These findings offer fresh insights into the mechanism of electronic friction and promising opportunities for friction control in device applications.
\end{abstract}




\clearpage
\newpage

Friction between surfaces often leads to significant mechanical energy dissipation, irreversible material deformation or degradation, and even mass loss, in both macroscopic and microscopic systems (\zfig{Fig. 1a})~\cite{rabinowicz2008friction}.
The friction force measures the rate at which applied work is expended, and can be minimized on ultra-flat and weakly interacting surfaces, where structural features of the contacts remain intact during sliding or rotation.
This scenario has recently been observed in systems of 2D crystals, where the corrugation of free energy surfaces with respect to interfacial motion is restricted.
At misaligned contact between mismatched or twisted lattices, the limit of structural superlubricity (SSL) can be achieved.
This presents the potential for creating devices with no wear and negligible friction or energy loss.
In this appealing context, kinetic energy is dissipated exclusively through the suprressed phononic and intrinsic electronic excitation and dissipation~\cite{hod2018structural,wang2024colloquium}.
Understanding physics behind electronic friction and exerting rational control of it thus holds great promise for designing high-performance, energy-efficient device applications with unlimited lifespans.

Phononic excitation and dissipation have been extensively studied within the framework of lattice dynamics theories.
Classical models, such as the Prandtl-Tomlinson (particle-on-chain) and Frenkel-Kontorova (chain-on-chain) models, provide intuitive insights and predictive capabilities~\cite{persson2013sliding}.
In contrast, the nature of electronic friction remains largely unclear for several reasons.
Experimentally, it is challenging to disentangle electronic friction from phononic contributions in the measured force and energy dissipation. 
To address this issue, frictional tests involving superconducting materials have been conducted, exploring temperature variations across the superconducting transition temperature~\cite{kisiel2011suppression,wang2020single}.
However, the superconducting transition in the bulk material does not directly elucidate the excitation processes at the interfaces, where friction primarily occurs~\cite{altfeder2012temperature,highland2006superconductivity}.
Direct characterization of electronic excitations at embedded and moving interfaces between material surfaces is challenging.
However, if we could manipulate the interfacial states, viable approaches might become available.

From the theoretical perspective, the key processes underlying phononic and electronic friction span a wide range of spatiotemporal scales (\zfig{Fig. 1b}).
Electron-hole excitation occurs at the femtosecond scale, while collective excitations, such as plasmons, can last up to picoseconds, comparable to phononic excitations~\cite{he2023current}.
Specifically, the washboard frequency $v/a$ driven by sliding at a speed $v$ over a lattice could reach the picosecond scale, where $a$ is the lattice constant or the periodicity of potential energy surface (PES) corrugation along the sliding path.
Mechanical movements, such as translation or rotation, typically occur at slow velocities (for instance, $\sim 1$~\textmu m $- 1$ m/s).
Nevertheless, nanoscale devices can achieve speeds ranging from $10^{2}$ to $10^{3}$ m/s~\cite{yang2013observation,peng2020load}.
This wide range contributes to the extensive spectrum of mechanical reconfiguration, encompassing electron diffusion rates of $\sim 0.01-1$ m/s in typical semiconductor materials~\cite{sverdrup2001sub-continuum,arab2016dopping}.
Recent studies demonstrate that the moving zone of dielectric depletion near the edges of a finite contact generates direct current, signaling mechanoeletrical energy conversion~\cite{huang2024tuning}.
Moreover, the overlap in the energy spectra of lattice vibration (phonons) in a solid or dielectric excitations (`hydrons') in a liquid and electronic excitations, such as plasmons, facilitates an interaction between these fundamental subsystems.
These principles are integrated into quantum friction theories within the framework of non-equilibrium fields~\cite{kavokine2022fluctuation}.
The entanglement between ionic and electronic degrees of freedom thus presents an intricate and promising area for further exploration.
It requires achieving control over the coupling process and being able to quantitatively estimate the contributions from various mechanisms.

By harnessing the atomically smooth and clean surfaces of 2D or van der Waals (vdW) crystals, we constructed an SSL platform device aimed at reducing phononic energy dissipation and elucidating the underlying mechanisms of electronic friction.
Using electrical and mechanical controls, we tuned the electronic coupling and charge density, directly measuring friction responses.
Our experiments and theory reveal that in an SSL system, electronic friction prevails over phononic contributions.
Notably, closing the vdW gap can eliminate electronic friction, which was experimentally achieved through either electrical or mechanical controls.
These findings provide key insights into electronic friction at the SSL limit of `ideal' interfaces and offer practical strategies for friction control in device applications.

\subsection*{Evidence of electronic friction at SSL contacts}
To elucidate the mechanisms of electronic friction, we create SSL `slidevices' from a vdW contact between graphite mesas (the `slider') and several 2D crystals (the `substrate') including molybdenum disulfide or MoS$_2$ (semiconducting), graphite (metallic), and hexagonal boron nitride or h-BN (insulating).
The mesa can slide over the substrate under a normal load in the device.
The friction force and current generation~\cite{huang2021microscale} across the contact can be measured during the sliding.
Micro-sized 2D substrate flakes were exfoliated from bulk 2H MoS$_{2}$ crystals (provided by HQ Graphene) and placed on a Si stage covered by an oxide layer.
Graphite mesa etched from HOPG was then transferred onto the flat regions (the root mean square or RMS is below $200$ pm) of the substrate flakes to form atomistically smooth 2D SSL contacts (\zfig{\textbf{Supplementary Note S1}}).
A normal force ($F\rm _N$) can be applied to the graphite mesa that slides on MoS$_{2}$ by using an atomic force microscopy (AFM) tip in atmospheric conditions(\zfig{\textbf{Supplementary Note S2}}).

This setup closes mechanical energy dissipation channels such as irreversible deformation, adhesion, and wear (\zfig{Fig. 1a}).
The incommensurability between graphite and other 2D crystals (lattice mismatch with h-BN and MoS$_{2}$, or twist between graphite) further suppresses the lattice-scale corrugation of interfacial PES.
Only intrinsic dissipation through weak phononic and electronic excitations can be activated.
To explore electronic excitation across the 2D contact, gate ($V_{\rm g}$) and bias ($V_{\rm b}$) voltage controls are integrated into the system.
The normal force ($F_{\rm N}$) is also controlled.
Friction characteristics are then probed via four parameters: the energy gap or electrical conductivity of substrate materials, $F\rm _N$, $V_{\rm g}$, and $V_{\rm b}$(\zfig{Fig. 1b}).

We first confirm the SSL state of the graphite-MoS$_{2}$ contact.
Normal forces are applied from $F_{\rm N} = 6$ to $66.5$ \textmu N step by step with an interval of $5.5$ \textmu N.
$20$ continuous sliding cycles were performed at each pressure level.
The sliding motion is controlled with an amplitude of $2$ \textmu m per cycle, at a speed of $4$ \textmu m/s.
The measurements show negligible dependence on the speeds in the range of $1-10$ \textmu m/s.
The friction force $f = \Delta E/\Delta s$ measured in the experiments, where $\Delta E$ is the mechanical energy dissipated along a sliding distance of $s$, which increases with $F_{\rm N}$ below a critical force $F_{\rm c} = 50$ \textmu N but decreases with $F_{\rm N}$ beyond $F_{\rm c}$ (\zfig{Fig. 1c}, $V_{\rm b} = V_{\rm g} = 0$ V).
Molecular dynamics (MD) simulations of the graphene-MoS$_2$ contact show that phononic friction remains invariant under normal pressure with differential coefficient of friction (DCOF) $< 10^{-5}$ (\zfig{Fig. 1c} and \zfig{\textbf{Supplementary Note S3}}).
The pressure dependence observed in the SSL regime cannot be explained in the picture of phonon-mediated friction.
Instead, beyond a critical pressure, the vdW 
gap between 2D materials can be closed, and electrons can flow across the interface without experiencing a barrier~\cite{jinbo2023vertical}.
We thus speculate that the electronic excitation at the graphite-MoS$_{2}$ heterojunction and subsequent dissipation in the semiconducting MoS$_{2}$ dominates the electronic friction.

The sliding speed in our experiments, on the order of $1$~\textmu m/s, is much lower than the diffusion velocity of electrons and holes in MoS$_{2}$~\cite{huang2024tuning}.
The moderate conductivity of charge carriers allows for notable electronic friction contributions during the relaxation process.
This theoretical discussion extends to the graphite-graphite and graphite-h-BN contacts.
The high electrical conductivity of graphite allows fast relaxation of electronic excitation and does not lead to significant energy dissipation.
At the h-BN contact, electronic excitation is prohibited by its wide band gap of $6$ eV~\cite{darshana2018monolayer}.
These estimates correspond exactly with our experimental results.

The friction force can then be decomposed into electronic and phononic contributions, where the latter can be quantified in the condition that the vdW gap is closed, corresponding to a low frictional stress of $14$ kPa in our SSL graphite-MoS$_{2}$ device (\zfig{Fig. 1d}).
The phononic friction force quantitatively agrees with those measured in the graphite-graphite (metal) and graphite-h-BN (insulator) systems, which is largely contributed by the edges ($\sim 75$\%, see \zfig{\textbf{Supplementary Note S4}} for details), and can be further suppressed to nearly zero by minimizing edge contact~\cite{li2024towards}.
This result suggests that the electronic friction is only significant at the metal-semiconductor interface with a notable vdW gap, and the electronic frictional stress is approximately $10-20$ kPa for the graphite-MoS$_{2}$ contact.

\subsection*{Electrical control of friction}
To gain more insights into the excitation and dissipation processes in friction, electrical controls are introduced.
The work functions of graphite and MoS\textsubscript{2} were measured as $5.05$ eV and $5.12$ eV, respectively, by using scanning Kelvin probe microscopy (SKPM) (\zfig{\textbf{Supplementary Note S5}}).
We begin by tuning the charge carrier density on the MoS$_{2}$ side of the contact by applying a gate voltage, $V_{\rm g}$, while maintaining a constant bias of $V_{\rm b} = 1$ V and a normal load of $F_{\rm N} = 36$ \textmu N.
The frictional force increases with $V_{\rm g}$ for a contact between graphite and single-layer MoS$_{2}$, suggesting that the energy dissipation via electronic excitation increases with more electrons at the interface (\zfig{Fig. 2a}).
Emphasizing that MoS$_{2}$ is an n-type semiconductor with electrons as charge carriers, positive gating adds extra electrons in MoS$_{2}$ rather than in SiO$_{2}$~\cite{novoselov2006unconventional}, while negative gating decreases them.
In the range $0 \leq V_{\rm g} \leq 15$ V, the carrier density within MoS$_{2}$ rises with $\lvert V_{\rm g} \rvert$, achieving $\sim 3.23 \times 10^{13}$ cm$^{-2}$ at $V_{\rm g} = 15$ V, which enhances electronic friction and $f$.
Conversely, when $V_{\rm g}$ takes on negative values ($-10$ V $< V_{\rm g} < 0$ V), there is a gradual depletion of most electrons in the channel, resulting in a decrease in $f$ with $\lvert V_{\rm g} \rvert$ (\zfig{\textbf{Supplementary Note S6}}).
At substantial negative gate voltages ($V_{\rm g} \leq -10$ V), interface trap states capture the residual electrons, thus preventing total depletion and maintaining the carrier density at a minimal level, corresponding to a change in the gate dependence of $f_{\rm el}$~\cite{byungwook2023one}.
In contrast, for a multilayer MoS$_{2}$ (approximately $10$ nm thick), the amplitude of $f$ remains relatively unaffected by the gate voltage for charged carriers are confined in the bottom layers and cannot reach the graphite-MoS$_2$ contact (\zfig{Fig. 2b}).

The recently proposed mechanism of Schottky generator allows us to decipher the characteristics of electronic excitation at the graphite-MoS$_{2}$ contact.
The generated current across the moving Schottky contact is measured by fixing the graphite mesa to the Au-coated AFM tip (the diameter is $d = 30$ nm), and driving the MoS$_{2}$ substrate by a piezoelectric ceramics stage moving in a line mode.
A circuit is connected to the conductive AFM (C-AFM) probe to monitor current flow across the static and sliding contacts using a gold contact.
The gating effect is also witnessed in the dependence of Schottky current generated from the sliding of SSL contact.
The insets in \zfig{Fig. 2a} and \zfig{2b} illustrate the current variation across $V_{\rm g}$ in the range of $\pm 15 \, \mathrm{V}$.
In the case of monolayer MoS$_2$, the current undergoes a variation spanning $6$ orders of magnitude, whereas for multilayer MoS$_2$, the current changes by just $2$ orders of magnitude, validating our understanding of the gating effect with a carrier concentration dependence.

To directly examine the effect of vdW gap closure, we tune the bias voltage, $V_{\rm b}$, and measure the friction force at $V_{\rm g} = 0$ and $F_{\rm N} = 36$ \textmu N.
The value of $f$ is constant for $V_{\rm b}$ between $E_{\rm g} = \pm 8$ V (\zfig{Fig. 2c}).
The electronic gap is closed out of this window, resulting in a significant reduction in $f$, which agrees well with the mechanically triggered transition (\zfig{Fig. 1c}).
The asymmetric bias effect at $\lvert V_{\rm b} \rvert > 8$ V is due to a $V_{\rm b}$-dependence of carrier concentration \cite{huang2024tuning}.
It should also be noted that, the vdW gap between graphite and MoS$_2$ ($E_{\rm b}$) is $\sim 4$ eV~\cite{jinbo2023vertical}, and by considering Kirchhoff's rule of voltage division, the potential drop at the interface, $V_{\rm d} = 4$ V, matches perfectly the measured value of $E_{\rm b} = 4$ eV (\zfig{\textbf{Supplementary Notes S5}}).
The stability of bias control is further demonstrated over the course of $50$ sliding cycles (\zfig{Fig. 2d}).
The swift and reversible behaviors of electronic friction present practical significance, with the timescale set by bias voltage application to close the vdW gap.

The capacitor model can be used to understand the gating and bias effects on electronic friction.
In a capacitor, when a voltage is applied between two conductive plates, charges accumulate on the plates. This charge accumulation effect can be applied to our monolayer MoS$_{2}$ samples.
When a gate voltage is applied, it effectively charges the MoS$_{2}$ layer, similar to one electrode of a capacitor, with the substrate acting as the other electrode.
The vdW gap is analogous to the dielectric material in a capacitor.
As the gate voltage increases, more carriers are introduced into the MoS$_{2}$ layer, enhancing its conductivity and altering its electronic states. This increase in carrier density significantly elevates $f$, as observed in our experiments, where the amplitude of $f$ is proportional to the carrier concentration.
On the other hand, when the bias voltage or pressure exceeds a certain threshold ($V_{\rm c} = 8$ V or $F_{\rm c} = 50$ \textmu N), the vdW gap is closed, leading to a significant reduction in friction.
This phenomenon can be explained by the suddenly enhanced conductivity across the interface and reduced electronic friction as the result, similar to the breakdown of dielectric material in a capacitor under high voltage.

\subsection*{Mechanical control of friction}
Mechanical control via normal loads can modulate both phononic and electronic friction.
However, our MD simulations and experimental data for graphite-graphite and graphite-MoS$_{2}$ contacts suggest a negligible pressure effect on phononic friction (\zfig{Fig. 1c} and \zfig{\textbf{Supplementary Note S3}}).
Moreover, the piezoelectric effects was excluded with a piezoelectric voltage of $178$ mV (\zfig{\textbf{Supplementary Note S7}}) which is  significantly smaller than the bias voltage applied in our experiments (typically $8$ V).
To probe electronic coupling and excitation at the 2D contact, we use C-AFM to measure the interfacial electrical conductance under the normal load.
Two pivotal values of the normal force ($F_{\rm s}$, $F_{\rm c}$) have been discerned from the observed friction force, marking the shift from a fluctuating regime to stages of increase and decrease \zfig{Fig. 3}.
Before the applied pressure reaches $F_{\rm c}$, the resistance remains very high, at the order of $10^7$ Ohms, and electron flow across the interface is strongly depressed.
However, once the pressure exceeds $F_{\rm c}$, the electron clouds at the contact overlap, closing the vdW gap and activating the electronic coupling. 
As a result, a sharp increase in conductance and a significant decrease in resistance to the order of $10^5$ Ohms are identified (\zfig{\textbf{Supplementary Note S8}}).

The transitions identified in the critical friction forces, $F_{\rm s}$ and $F_{\rm c}$, align with those in the generated current density $J$ (\zfig{Fig. 3a}).
Below $F_{\rm s} = 28$ \textmu N, a noisy current from $-2$ A/m$^{2}$ to $-15$ A/m$^{2}$ is observed.
The DCOF increased from nearly zero ($ < 10^{-5}$) to $0.01$.
The average value is $J = -3$ A/m\textsuperscript{2}.
This observation is explained by the edge contact that is firstly enhanced by the pressure and then weakened due to the warp effect~\cite{zhanghui2024positive-negative,hu2024effects}.
The generated current is attributed to the reconstruction of depletion zones in the semiconductors~\cite{huang2021microscale}.
Beyond $F_{\rm s}$, the edge contact is lost and the generated current suddenly diminishes with a noise level below $0.01$ A/m$^{2}$(\zfig{Fig. 3b}).
A full 2D contact is thus formed, and the electronic friction increases with the pressure till the normal force reaches $F_{\rm c} = 50$ \textmu N, where the generated current density surged to $J = 5$ A/m\textsuperscript{2}~\cite{kang2014computational}.
This pressure dependence of $f$ in the range of $F_{\rm s} < F_{\rm N} < F_{\rm c}$ originates from the charge transfer across the graphite-MoS$_{2}$ interface that increases with the amplitude of load, $F_{\rm N}$~\cite{jinbo2023vertical}.
The amplitude of electronic friction scales with the area of the mesa and thus is significantly higher than that due to the edge contact.
Further increase of the normal force beyond $F_{2}$ closes the vdW gap, analogous to the the situation with a bias voltage beyond the threshold defined by the band gap.

\subsection*{Quantitative understanding of electronic friction}
Electronic friction observed at the graphite-MoS$_{2}$ contact in our experiments could arise from the injected charge of moving depletion zone or from quantum fluctuation-induced within MoS$_{2}$ and graphite\cite{volokitin2011quantum, kavokine2022fluctuation}.
We first evaluate the electronic friction rising from doping-induced charge carriers in MoS$_{2}$ at the moving contact, where the charges are considered to be static relative to MoS$_{2}$ and the surface response function of graphite is described with the Drude model\cite{kavokine2022fluctuation}.
The analysis reports an electronic friction stress falling into the range between $0.8$ Pa and $1.6\times 10^{-2}$ Pa, depending on the assumed separation between charges in MoS$_2$ and the graphite surface.
Moreover, the friction induced by charge transfer-induced interfacial dipoles at the MoS$_{2}$-graphite interface is estimated to be $4.7 \times 10^{-3}$ Pa, and the additional contributions from fluctuating charges and phonons in MoS$_{2}$ and graphite are also significantly lower than the values measured in our sliding contact experiments (\zfig{\textbf{Supplementary Note S9}}).

Next, to explore the dominant contribution from moving depletion zones at the finite-sized contact, perturbative finite element analysis (pFEA) was conducted (see \zfig{Methods} for details).
Since the sliding speed in our experiments is much slower than electron and thermal diffusion in MoS$_{2}$ ($\sim 1$ and $0.002$ m/s, respectively~\cite{sverdrup2001sub-continuum,takamura1987thermal}), frictional energy dissipation is governed by electronic excitations at the moving contact.
The subsequent redistribution of carriers in MoS$_{2}$ leads to energy dissipation through electronic friction.
\zfig{Fig. 4a} and \zfig{4b} exhibit the potential distribution over time, showing electron displacement as a Schottky contact is reestablished as graphite moves continuously on MoS$_2$.
The direction of the electric field or current flow indicates that the displaced electrons slowly achieve equilibrium through diffusion, which is characterized by a time constant of $\tau = 1.32$ ps (\zfig{\textbf{Supplementary Note S10}}).
This is consistent with the experimentally observed direction of generated current at $F_{\rm N} < F_{\rm s}$ (\zfig{Fig. 3a}).

The electronic friction or energy dissipation is quantified by $f_{\rm el} = {\rm d}q/{\rm d}s  = \Delta E/\Delta s$, where $q$ is the energy dissipated via Joule heat and other scattering processes within the semiconductor, implicitly considered in modeling, and $E = \frac{1}{2} \iiint \mathbf{D} \cdot \mathbf{E} \, dV$ is the incremental electrostatic energy at time $t$, along with the sliding distance $\Delta s$.
$\mathbf{D}$ and $\mathbf{E}$ are the electric displacement vector and electric field, respectively.
When the normal force $F_{\rm N}$ is less than $F_{\rm c}$ (\zfig{Fig. 4c}), increasing this force triggers interfacial charge transfer~\cite{jinbo2023vertical}, which elevates the charge density at the interface, thus enhancing energy dissipation.
Conversely, when $F_{\rm N}$ exceeds $F_{\rm c}$, the interface becomes conductive, aligning the potentials of graphite and MoS$_2$, which reduces electronic friction.
Augmenting the gate voltage leads to increased dissipation by elevating the electron concentration within MoS$_2$ (\zfig{Fig. 4d}).
At $V_{\text{g}} < -10$ V, electrons are nearly depleted, and more negative gate voltage decreases electron concentration significantly, resulting in a strong $V_{\rm g}$-dependence of electronic friction.
Additionally, when $V_{\text{b}}$ exceeds $\pm 8$ V, the vdW gap closes, akin to the condition of $F_{\rm N} > F_2$, equalizing the chemical potentials of electrons in graphite and MoS$_2$, thereby mitigating electronic friction (\zfig{Fig. 4e}).
These observations align well with our experimental data for the graphite-MoS$_{2}$ contact, and the model setup apparently indicates the absence of electronic friction in the metallic (graphite-graphite, no chemical potential difference across the interface) and insulating (graphite-h-BN, no charge carriers) SSL contacts ({\zfig{\textbf{Supplementary Note S11}}}).


\section*{Conclusion}\label{sec5}
Based on van der Waals device setups, we demonstrate the disentanglement and control of electronic friction from phononic contributions via electrical and mechanical control.
The findings reveal the pathway of energy dissipation as a result of electronic excitation at the moving contact with a constantly reconstructing charge depletion zone, and subsequent relaxation in the semiconducting materials.
Regulated electrical friction can achieve magnitudes comparable to phonon friction and exhibits a pronounced dependence on the electronic structures of materials at the interface, which, however, can be eliminated by closing the van der Waals gap.
Vanishing electronic friction can also be achieved at metal/metal, metal/insulator contacts, which can also be made structurally superlubric for applications with no wear and minimal friction for unlimited durability.
These findings provide new avenues for manipulating friction in the on-device settings, opening unconventional opportunities in microsystem design.

\clearpage
\newpage

\section*{Models and Methods}\label{sec2}

\subsection*{Construction of SSL contacts}
Substrate materials possessing single-crystalline surfaces, including 2D hexagonal boron nitride (h-BN), 2D molybdenum disulfide (MoS$_{2}$), and graphite, were exfoliated from their bulk forms.
Subsequently, micrometer-sized graphite flakes (the `mesa') cut from highly oriented pyrolytic graphite (HOPG) and exhibiting a single-crystalline surface were placed onto the substrate with mismatched lattice constants (h-BN and MoS$_{2}$) or a twisted interface (graphite), thereby establishing the SSL contact as verified by friction characteristics measurements (\zfig{\textbf{Supplementary Note S1}})~\cite{yu2024on-device}.

The representative Schottky junction between MoS$_2$ and graphite shows an ideality factor varying between $1.03$ and $1.07$~\cite{yu2024on-device}, signifying behavior that is close to an ideal diode.
As the contact forms, electronic excitation processes push charge carriers (holes for MoS$_2$) toward graphite, yet a van der Waals (vdW) gap impedes this movement.
The built-in electric field of the depletion region counteracts this charge transfer to preserve equilibrium.
The Schottky barrier height, $\phi = 0.6$ V, determines the energy barrier for charge transfer, which is crucial for the electronic properties of the junction and modulating friction at the interface via electric control.
In mechanical control, the contact resistance changes with the applied pressure, which is $\sim 10^{7}$ Ohm below a pressure threshold below $P_{\rm c} = 50$ \textmu N, and decreases to $10^{5}$ Ohm beyond $P_{\rm c}$, where the vdW gap is closed.

\subsection*{Control of electronic friction}
In electrical control, gate and bias voltages were applied to modulate the electronic states at the contact interfaces.
For gate control, a back gate is introduced by using monolayer and multilayer MoS$_2$ samples prepared on SiO$_{2}$/Si substrates.
$V_{\rm g}$ varies from $-15$ V to $15$ V with a scanning speed of $0.06$ V/s, and the measurement is conducted at $V_{\rm b} = 1$ V, $F_{\rm N} = 36$ \textmu N.
For bias control, a series of voltage from $V_{\rm b} = -10$ to $10$ V are applied via an Au-capped probe, increasing by $2$ V each step.
$50$ continuous sliding cycles were performed at each level of bias voltages.

In mechanical control, a normal force was applied to the system by an atomic force microscopy (AFM) tip to achieve high local pressure.
A series of normal forces ranging from $6$ \textmu N to $66.5$ \textmu N are applied via an Au-capped probe, step by step, with an interval of $5.5$ \textmu N.

\subsection*{First-principles calculations}

Density functional theory (DFT)-based first-principles calculations were conducted to characterize the contact physics using the Vienna \textit{Ab initio} Simulation Package (VASP).
The exchange-correlation functional was described using the generalized gradient approximation (GGA) with the Perdew-Burke-Ernzerhof (PBE) parameterization.
The plane-wave cutoff energy is $520$ eV.
To eliminate interactions between periodic supercells, a vacuum layer of $0.2$ nm at $c$  direction was introduced.
The Brillouin-zone integration was performed using the Monkhorst-Pack k-point grid with a mesh density of $1 \times 2 \times 1$ for the supercell.
The computational supercell consisted of one layer of an armchair-edged graphene ribbon ($5 \times 5$) and one layer of 2D MoS$_2$ ($4 \times 12$) without edges, by enforcing periodic boundary conditions (PBCs), presenting the $x$ direction lattice mismatch around $1.2$\%.


\subsection*{Perturbative finite element analysis (pFEA)}
The pFEA setup employs a model characterized by a top slider with dimensions of length $L_1 = 4$ \textmu m and height $h_1 = 50$ nm to represent graphite, alongside a bottom stator with dimensions of length $L_2 = 10$ \textmu m and height $h_2 = 0.65$ nm to represent a monolayer of MoS$_2$~\cite{huang2024tuning}.
The doping level in MoS$_{2}$ is $N_{\rm A} = 10^{18}$ cm$^{-3}$.
Graphite is analogously represented by a heavily doped n$^{+}$-Si, with a doping concentration of $N_{\rm D} = 2 \times 10^{20}$ cm$^{-3}$, the changes in which in the highly-doped regime do not modify the results.
A continuity condition of electric displacements is enforced at the heterojunction interface.
Metallic contacts are established at the top of the slider and the bottom of the stator, maintained at potentials of $0$ and $V_{\rm b}$, respectively, resulting in an applied bias voltage of $V_{\rm b}$.

Within the semiconductor, the potential distribution adheres to the Poisson equation.
The carrier concentration is derived from the electronic band structures using the Fermi-Dirac statistics.
By resolving these equations, one obtains the potential and carrier concentration profiles, enabling the calculation of electron and hole currents via the drift-diffusion model using numerical solvers~\cite{selberherr1984analysis}.

FEA proceeded through three distinct phases.
Initially, in Step $1$, the distributions of electrons and holes within the Schottky diode were evaluated to ensure static equilibrium ($\Delta s = 0$) under various conditions.
Following equilibrium establishment, Step $2$ involved moving the upper slider by a perturbation, $\Delta s = 0.5$, while the carrier distributions in both the slider and stator were kept constant, thereby inducing a non-equilibrium state.
In Step $3$, the previously applied constraint was lifted, and a transient simulation commenced, using the non-equilibrium state from Step $2$ as its initial condition.
This stage analyzed the dynamics of carrier transport, alongside the electric potential distribution and associated parameters including the electric field, current density, band structure, and electrostatic energy over time.

\clearpage
\newpage


\section*{Acknowledgements}
This study was supported by the National Natural Science Foundation of China through grants 12425201 and 52090032.
The computation was performed on the Explorer 1000 cluster system of the Tsinghua National Laboratory for Information Science and Technology.

\subsection*{Conflict of interest}
The authors declare no competing interest.


\clearpage
\newpage


\begin{figure}[htp]
\centering
\includegraphics[width=1\linewidth]{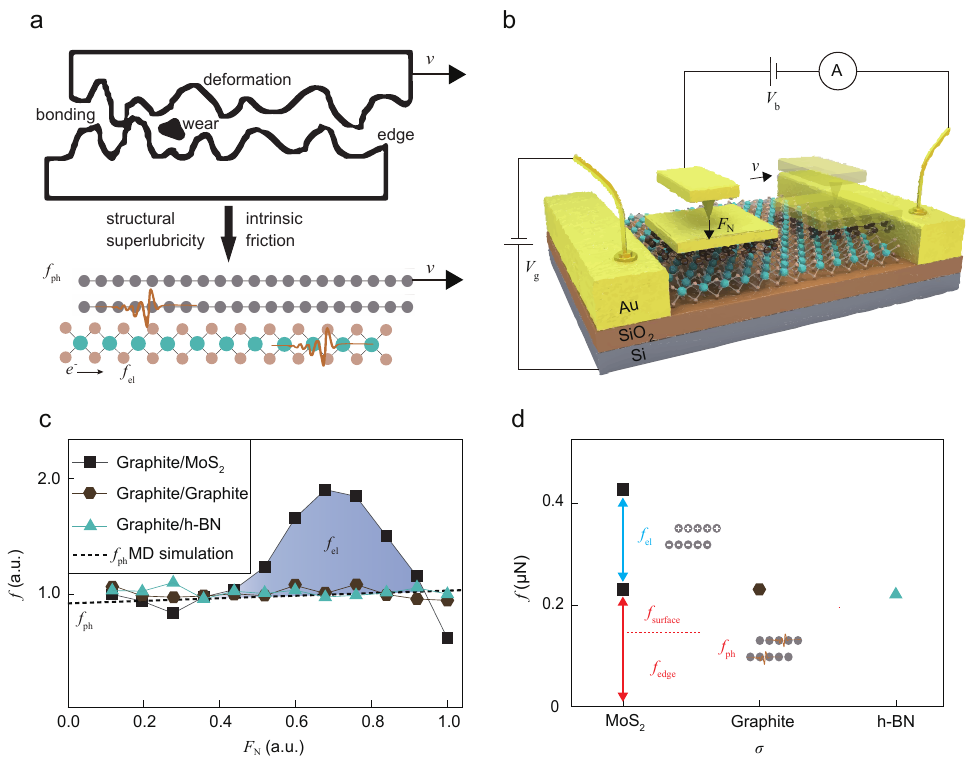}
\caption{Friction in sliding devices.
(a) Friction often involves a rich spectrum of microscopic mechanisms including bond breakage and reconstruction, adhesion, irreversible material deformation, damage and wear, and substantial edge effects, which can all be suppressed by constructing a structural superlucitiy (SSL) contact.
(b) The sliding device (`slidevice') setup features a moving SSL interface to minimize the number of energy dissipation channels to only the intrinsic phonons and electronic ones, allowing for precise mechanical and electrical control.
The setup consists of a heterojunction between micrometer-sized graphite and 2D MoS$_{2}$ (or graphite, 2D h-BN), placed on a SiO$_{2}$/Si substrate.
(c) The frictional force ($f$) plotted as a function of normal force ($F_{\rm N}$) at the contact between graphite and graphite (metal), 2D MoS$_2$ (semiconductor), and 2D h-BN (insulator) under mechanical control.
The results show that the electronic friction channel is closed with $F_{\rm N}$ above a threshold at the metal-semiconductor contact.
Molecular dynamics (MD) simulations suggest that phonon contribution ($f_{\rm ph}$) remains as constant as $f_{\rm N}$ varies, exhibiting a differential coefficient of friction (DCOF) $< 10^{-5}$.
(d) Phononic ($f_{\rm ph}$) and electronic ($f_{\rm el}$) contributions to friction force, where the residual phonon friction is expected to be predominantly arising from edge contacts with the micrometer-sized graphite. Using the equation $f = \tau L^2 + \gamma L$ where $L$ is the edge length, the edge friction coefficient \(\gamma = 8.4 \pm 0.6 \, \text{nN/\textmu  m}\) and bulk frictional stress \(\tau = 25.87 \pm 0.08 \, \text{kPa}\) are obtained~\cite{li2024towards}.}
\end{figure}

\clearpage
\newpage

\begin{figure}[htp]
\centering
\includegraphics[width=1\linewidth]{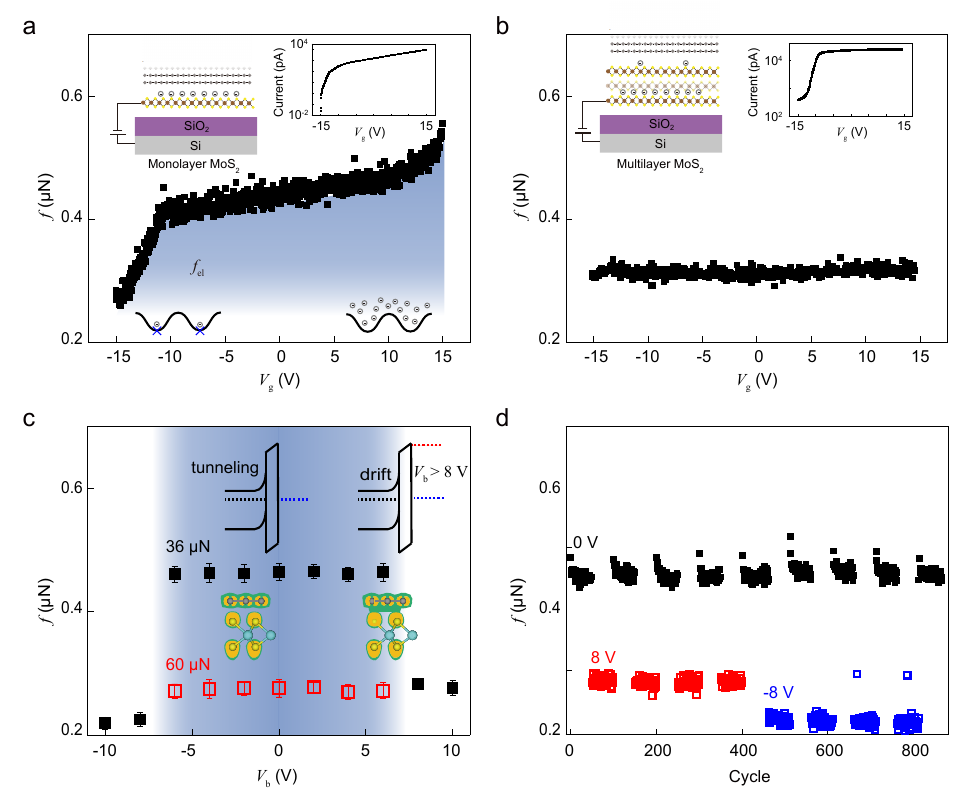}
\caption{Gate and bias control of electronic friction.
(a) Friction force ($f$) plotted as a function of gate voltage ($V_{\rm g}$) for the graphite-2D MoS$_2$ heterojuction, under zero bias voltage.
The bipolar response of $f$ to $V_{\rm g}$ is attributed to the modulation of charge carrier density at the contact.
(b) $f-V_{\rm g}$ relationship for the heterojunction between graphite and multilayer ($17.6$ nm) MoS$_2$.
The friction force remains relatively unaffected since the gate-controlled charge carrier density is confined in the bottom layer and cannot reach the heterointerface with graphite.
(c) $f-V_{\rm b}$ relationship for the graphite-2D MoS$_2$ heterojuction at zero gate voltage and normal forces of $F_{\rm N} = 36$ \textmu N (with an vdW gap) and $60$ \textmu N (vdW gap closed), respectively.
A threshold voltage of $\pm8$ V is identified at $F_{\rm N} = 36$ \textmu N, above which the electronic friction is nearly eliminated.
The electronic channel of friction remains inactive as the vdW gap is closed at $F_{\rm N} = 60$ \textmu N.
(d) Robustness of bias control.
The bias voltage ($V_{\rm b}$) was cycled continuously between $-8$ V, $0$ V, and $8$ V, with each voltage maintained for approximately $50$ sliding cycles.
The result highlights the reversibility, consistency, and rapid response characteristics of bias control.}
\end{figure}

\clearpage
\newpage

\begin{figure}[htp]
\centering
\includegraphics[width=1\linewidth]{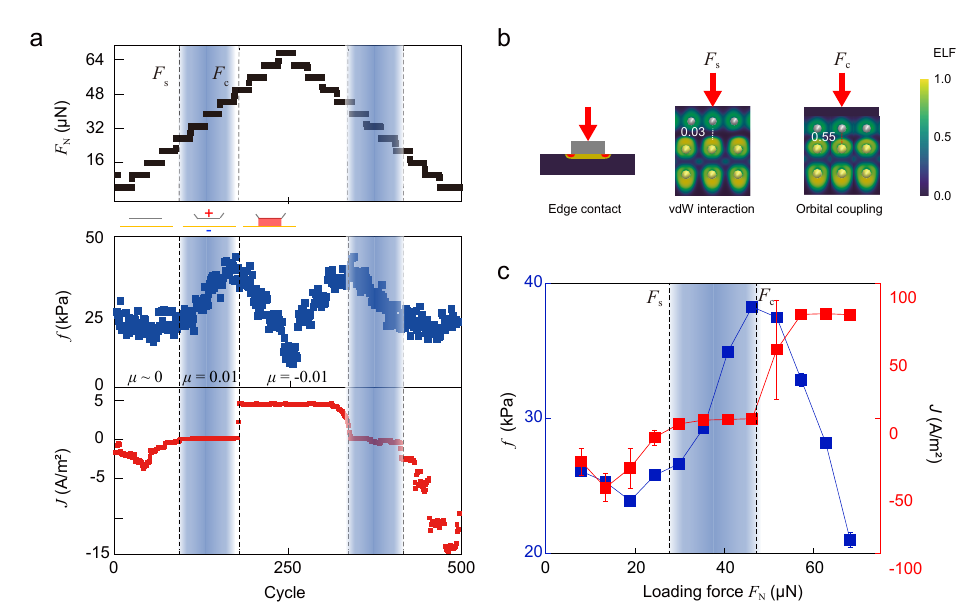}
\caption{Pressure control of friction.
(a) The friction force ($f$) and current ($J$) generated by the sliding Schottky contact as a function of normal force ($F_{\rm N}$)~\cite{yu2024on-device}.
Measurements were conducted with a series of normal forces ranging from $F_{\rm N} = 6$ \textmu N to $66.5$ \textmu N, with an interval of $5.5$ \textmu N.
Two threshold pressure values are noted as $F_{\rm s}$ and $F_{\rm c}$, which indicate the formation of full SSL contact and the start of vdW gap closure, which are illustrated in panel (b).
(c) Averaged value of $f$ and $J$ in each constant-force scan, showing two transitions that correspond to $F_{\rm s}$ and $F_{\rm c}$, respectively.}
\end{figure}

\clearpage
\newpage

\begin{figure*}[htp]
{
\includegraphics[width=\textwidth]{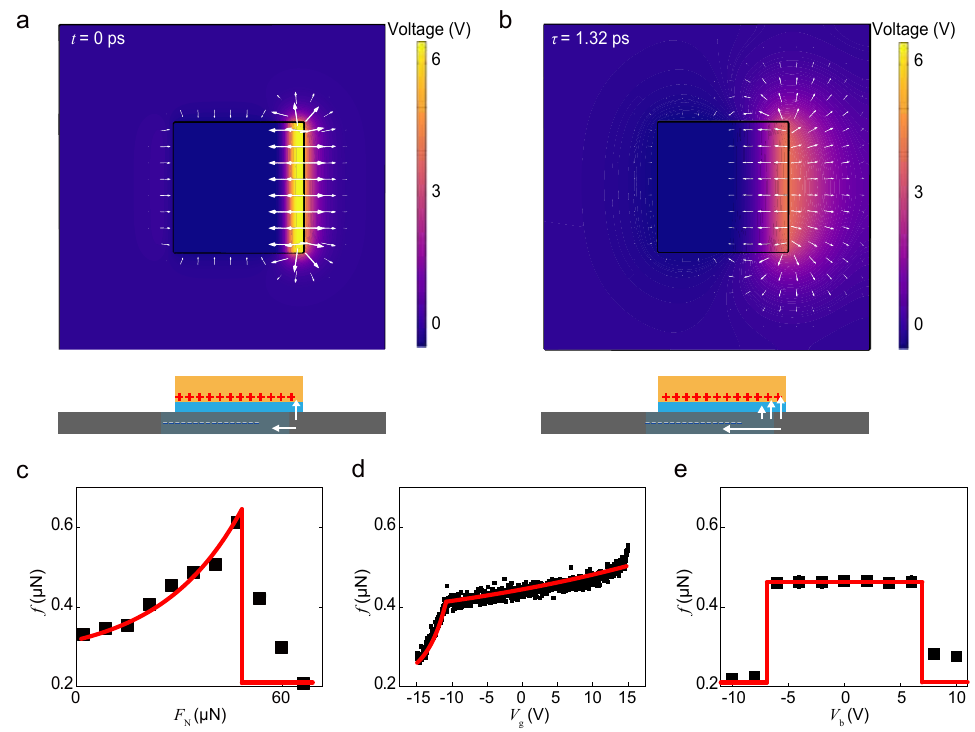}
\caption{Mechanism of electronic friction control.
Electrical potential responses to sliding motion of the SSL contact at (a) $t = 0$ ps and (b) $t = \tau = 1.32$ ps, exhibiting displacement and relaxation of the depletion zone in the basal planes and driving current generation across the interface.
The lines are results obtained from perturbative finite element analysis (perturbative finite element analysis or pFEA, see \zfig{Methods} for details), which well reproduce experimental findings of mechanical and electrical control for (c) normal force ($F_{\rm N}$), (d) gate voltage ($V_{\rm g}$), and (e) bias voltage ($V_{\rm b}$).}
}
\end{figure*}

\clearpage
\newpage

\bibliography{main_text}


\end{document}